\documentclass[12pt]{amsart}

\usepackage{amssymb}
\usepackage{epsfig}
\usepackage{comment}
\usepackage{amsmath}
\usepackage{caption}
\usepackage[section]{placeins}
\usepackage{graphicx}
\usepackage{units}
\usepackage{color}
\usepackage{subfig}
\usepackage{threeparttable}
\usepackage{bm}
\usepackage{algorithm}
\usepackage[noend]{algpseudocode}
\usepackage[margin=1.0in]{geometry}
\usepackage{multirow}

\makeatletter
\def\BState{\State\hskip-\ALG@thistlm}
\makeatother

\graphicspath{ {Figures/} }

\theoremstyle{definition}

\numberwithin{equation}{section}
\theoremstyle{remark}

\begin{document}

\title[]{Size and Shape Dependence of Hydrogen-Induced Phase Transformation and Sorption Hysteresis in Palladium Nanoparticles}

\author[]
{Xingsheng~Sun${}^\dagger$, Rong~Jin}

\address
{
    Department of Mechanical and Aerospace Engineering,
    University of Kentucky,
    Lexington, KY 40506, USA.
    ${}^\dagger$ Corresponding author. E-mail address: xingsheng.sun@uky.edu.
}

\begin{abstract}

We establish a computational framework to explore the atomic configuration of a metal-hydrogen (M-H) system when in equilibrium with a H environment. This approach combines Diffusive Molecular Dynamics with an iteration strategy, aiming to minimize the system's free energy and ensure uniform chemical potential across the system that matches that of the H environment. Applying this framework, we investigate H chemical potential-composition isotherms during the hydrogenation and dehydrogenation of palladium nanoparticles, ranging in size from $3.9$ nm to $15.6$ nm and featuring various shapes including cube, rhombic dodecahedron, octahedron, and sphere. Our findings reveal an abrupt phase transformation in all examined particles during both H loading and unloading processes, accompanied by a distinct hysteresis gap between absorption and desorption chemical potentials. Notably, as particle size increases, absorption chemical potential rises while desorption chemical potential declines, consequently widening the hysteresis gap across all shapes. Regarding shape effects, we observe that, at a given size, cubic particles exhibit the lowest absorption chemical potentials during H loading, whereas octahedral particles demonstrate the highest. Moreover, octahedral particles also exhibit the highest desorption chemical potentials during H unloading. These size and shape effects are elucidated by statistics of atomic volumetric strains resulting from specific facet orientations and inhomogeneous H distributions. Prior to phase transformation in absorption, a H-rich surface shell induces lattice expansion in the H-poor core, while before phase transformation in desorption, surface stress promotes lattice compression in the H-rich core. The magnitude of the volumetric strains correlates well with the size and shape dependence, underlining their pivotal role in the observed phenomena.

\end{abstract}

\maketitle


\section{Introduction}
\label{sec:intro}

Many energy- and information-storage processes depend on the phase transformation of nanomaterials within reactive environments. This transformation can be induced by solute intercalation and hence finds application in diverse fields such as hydrogen storage and detection~\cite{schlapbach2001hydrogen, wadell2014plasmonic}, metal hydride batteries~\cite{oumellal2008metal}, lithium-ion batteries~\cite{ebner2013visualization}, and nanoparticle synthesis~\cite{son2004cation}. Compared to their bulk counterparts, nanostructured materials are able to exhibit accelerated charging and discharging kinetics, increased energy and power density, extended life cycles, and tunable thermodynamics based on size and shape.

Metal hydrides, especially palladium hydride ($\text{PdH}_x$), serve as a convenient model system for studying solute-induced phase transformation, as they are characterized by fast hydrogenation and dehydrogenation kinetics at readily accessible temperatures and pressures~\cite{manchester1994h, flanagan1991palladium}. The phase diagram of bulk $\text{PdH}_x$ reveals that Pd initially forms a dilute interstitial solid solution with H, termed the $\alpha$ phase, wherein the lattice constant undergoes slight expansion with increasing H concentration. As H concentration further rises, e.g., beyond $x>0.015$ at room temperature, a phase transformation to the lattice-expanded $\beta$ phase takes place. Continued incorporation of H leads to further expansion of the $\beta$ phase lattice. The $\alpha$ phase eventually vanishes upon completion of the transformation, typically beyond $x>0.6$ at room temperature. Throughout these two phases, the face-centered cubic (FCC) structure is preserved within the Pd lattice. The octahedral site of the FCC lattice is energetically favored for H to occupy in both the $\alpha$ and $\beta$ phases.

While the first report of H absorption in Pd traces back to $1866$~\cite{graham1866xviii}, the exploration of the intrinsic, nanoscale size- and shape-dependent thermodynamic properties of this system is still in its infancy. Recent investigations into nanoscale systems have unveiled significant thermodynamic deviations from bulk behavior~\cite{zalineeva2017octahedral, langhammer2010size1}, likely due to their high surface-area-to-volume ratio. Given the inherent challenges in measuring H in Pd nanoparticles, there exists a wide array of experimental techniques and sample conditions. Currently, two primary approaches have been employed to probe the thermodynamic behaviors of nanosized $\text{PdH}_x$: ensemble and individual measurements.

Ensemble measurements entail studying the collective behaviors of numerous Pd nanoparticles interacting with H environments~\cite{langhammer2010size, narehood2009x, wadell2014thermodynamics, pundt2004hydrogen}. Consequently, these measurements are averaged across particles with potentially different sizes, shapes, and surface conditions, making it challenging to discern intrinsic size- and shape-dependent properties from effects caused by the morphological dispersity of samples. For instance, ensemble studies of Pd nanocubes ranging in size from $14$ to $65$~nm have shown sloped isotherms~\cite{bardhan2013uncovering}. Additionally, ensemble measurements of Pd nanoparticles with diameters of $2.6$ nm and $7$ nm also suggest a continuous transition from the H-poor $\alpha$ phase to the H-rich $\beta$ phase~\cite{yamauchi2008nanosize}. These findings present a significant contrast to the sharp $\alpha$-to-$\beta$ phase transformation observed in bulk and individual nanosized Pd-H systems and could profoundly impact materials design for energy and information storage applications. However, it still remains uncertain whether some of the observed trends result from intrinsic size effects or from particle dispersity inherent in ensemble measurements.

Individual measurements, on the other hand, offer a promising avenue to resolve this ambiguity by directly observing phase transformations in isolated nanoparticles~\cite{vadai2018situ, narayan2016reconstructing, hayee2018situ}. For instance, a sharp $\alpha$-to-$\beta$ phase transition and a size-dependent absorption plateau pressure have been unveiled in individual Pd nanocubes spanning lengths from $13$ to $29$~nm~\cite{baldi2014situ}. Size- and shape-independent enthalpies and entropies have been found for Pd nanoparticles ranging in size from $18$ to $63$~nm~\cite{syrenova2015hydride}. Furthermore, examination of pressure-composition isotherms for individual Pd nanocubes and nanooctahedra has shown that nanoparticles of these two shapes hydrogenate at similar pressures without equilibrium phase coexistence~\cite{sytwu2018visualizing}. Other notable findings from individual measurements include the observation of an atomistically sharp hydride phase boundary \cite{narayan2017direct}, the dynamics of avalanching strain during phase transformation \cite{ulvestad2015avalanching}, dislocation healing facilitated by phase transformation \cite{ulvestad2017self}, and the deceleration of phase transformation due to grain growth \cite{alekseeva2021grain}. However, these initial single-particle studies often rely on indirect physical phenomena of the two phases, such as electron energy-loss spectra~\cite{sytwu2018visualizing} and plasmon resonance~\cite{syrenova2015hydride}, to capture the phase transformation. Consequently, they may not fully capture the detailed distribution of H atoms and the resultant lattice deformation and distortion, thus limiting the depth of insight into the corresponding mechanisms.

In this study, we utilize an advanced computational technique known as Diffusive Molecular Dynamics (DMD) to explore the hydriding and dehydriding phase transformations in Pd nanoparticles of varying sizes and shapes. Our investigation aims to elucidate the impact of nanoparticle morphology on transformation chemical potential and sorption hysteresis. DMD represents a cutting-edge approach for simulating long-term diffusive mass transport and heat transfer while preserving atomic-level resolution~\cite{kulkarni2008variational, venturini2014atomistic, li2011diffusive, sun2016deformation}. This method integrates a discrete kinetic model governing the evolution of local solute concentrations at sites over diffusive timescales with a nonequilibrium statistical thermodynamics model that enables the relaxation of the crystal structure. The latter also provides the requisite driving forces for kinetics. In our work, we establish a computational framework by integrating DMD with an iteration strategy and apply it to nanosized Pd-H systems. This framework is able to determine the distribution of local H concentrations and the resultant lattice deformation while minimizing the system's free energy. In contrast to other atomistic methods such as transition state theory-based accelerated molecular dynamics (MD) approaches \cite{voter1998parallel, voter1997method, so2000temperature}, kinetic Monte Carlo (MC) methods \cite{henkelman2001long}, phase field crystal methods \cite{berry2008melting}, atomic density function methods \cite{jin2006atomic}, and quasi-particle approaches \cite{demange2022atomistic}, our proposed DMD-based framework offers the advantage of ensuring uniform chemical potential across the system. This feature allows for the maintenance of chemical equilibrium with a H environment, a setup that is more physically realistic compared to directly controlling the total number of H atoms in the system.

The remainder of this paper is organized as follows. Section~\ref{sec:method} offers a concise overview of DMD model equations and introduces the iteration strategy for determining equilibrium atomic configurations. Following this, Section~\ref{sec:result} presents the outcomes of our numerical experiments. We discuss absorption and desorption isotherms, plateau chemical potentials, hysteresis gaps, and the saturation ordering in hydride and dehydride phase transformations. To elucidate the observed size and shape effects, we calculate and analyze atomic volumetric strains at chemical potentials before phase transformation in both absorption and desorption. Finally, in Section~\ref{sec:concl}, we summarize our findings and present concluding remarks.

\section{Methodology}
\label{sec:method}

For the sake of convenience and completeness, this section begins with a brief overview of the DMD theory and its application in simulating H diffusion within metallic nanocrystals. Interested readers can delve deeper into the topic through references such as \cite{kulkarni2008variational, li2011diffusive, venturini2014atomistic, sun2017acceleration, simpson2016theoretical}. Following this, we introduce a computational framework aimed at determining the equilibrium configuration of H local concentrations and atomic positions during both absorption and desorption processes, while maintaining a fixed chemical potential. Subsequently, we present the setup of simulations and some implementation details in this work.

\subsection{A nonequilibrium thermodynamics model}

We focus on a three-dimensional metal-hydrogen (M-H) system characterized by two types of lattice sites: host and interstitial sites. We assume that metallic atoms always occupy host sites, while interstitial sites can accommodate either a H atom or remain vacant. We denote the sets of host and interstitial sites as $I_\text{M}$ and $I_\text{H}$, respectively. The total numbers of host and interstitial sites are represented by $N_\text{M}$ and $N_\text{H}$, respectively, i.e., $N_\text{M}=|I_\text{M}|$ and $N_\text{H}=|I_\text{H}|$. At each interstitial site $i\in I_\text{H}$, we define an occupancy function $n_i$ as~\cite{sun2019atomistic}
\begin{equation}
n_i =
 \begin{cases}
    1      & \text{if the site } i \text{ is occupied by a H atom},\\
    0      & \text{if the site } i \text{ is vacant}.\\
  \end{cases}
  \label{eq:occ}
\end{equation}
In contrast, the occupancy of the host sites remains constant at $1$, i.e., $n_i = 1$ for $i\in I_\text{M}$. From the definition in Eq.~(\ref{eq:occ}), it follows that the occupancy array $\{n\}$ can take values from a set comprising elements of $\{0,1\}$. This set, denoted by $\mathcal{O}$, is defined as
\begin{equation}
\mathcal{O} = \{0,1\}^{N_\text{H}}.
\end{equation}

Additionally, the instantaneous position and momentum of site $i$ are denoted by $\bm{q}_i$ and $\bm{p}_i$, respectively. Over time scales significantly longer than atomic vibrations, we assume that these microscopic state variables can be treated as random variables characterized by a joint probability density function (PDF) $\rho\big(\{\bm{q}\}, \{\bm{p}\}, \{n\}\big)$, where $\{\bm{q}\}=\{\bm{q}_{i}:i\in I_\text{M}\cup I_\text{H}\}$, $\{\bm{p}\}=\{\bm{p}_{i}:i\in I_\text{M}\cup I_\text{H}\}$ and $\{n\}=\{n_{i}:i\in I_\text{H}\}$~\cite{sun2019atomistic}. Consequently, we can define the expectation or macroscopic value of any quantity $A\big(\{\bm{q}\}, \{\bm{p}\}, \{n\}\big)$ using the well-established phase average in classical statistical mechanics~\cite{kulkarni2008variational}
\begin{equation} 
\langle A \rangle
= \sum_{\substack{\{n\}\in\mathcal{O}}}
\dfrac{1}{h^{3(N_\text{M}+N_\text{H})}}
\int A\big(\{\bm{q}\}, \{\bm{p}\}, \{n\}\big) \rho\big(\{\bm{q}\}, \{\bm{p}\}, \{n\}\big)
\prod_{i\in{I_\text{M}} \cup I_\text{H}}\text{d}\bm{q}_i\text{d}\bm{p}_i,
\label{eq:ave}
\end{equation}
where $h$ is the Planck's constant. Since the PDF is a highly complex function depending on nonlinear Hamiltonians, the analytical solution of Eq.~(\ref{eq:ave}) in closed-form is generally intractable, leading to the need for approximation theory. {\it Venturini et al.}~\cite{venturini2014atomistic} proposed a mean-field theory, where the integration problem is addressed within a finite-dimensional trial space spanned by a predetermined class of trial Hamiltonians. Following this, we further assume that $\bm{q}_i$ and $\bm{p}_i$ of each host and interstitial sites follow normal distributions, whereas $n_i$ of each interstitial site follows a Bernoulli distribution~\cite{li2011diffusive, dontsova2014solute, dontsova2015solute}. Then the trial PDF $\rho_0$ can be expressed as~\cite{sun2017acceleration}
\begin{equation} 
\begin{aligned}
\rho_0 \big( \{\bm{q}\}, \{\bm{p}\}, \{n\} \big)  =
& \dfrac{1}{\mathnormal{\Xi}_0} \exp\Bigg(-\sum_{i\in{I_\text{M} \cup I_\text{H}}}
\bigg(\frac{1}{2\sigma_i^2}|\bm{q}_i - \bar{\bm{q}}_i|^2
+ \frac{1}{2k_{\text{B}}T_i m_i}|\bm{p}_i - \bar{\bm{p}}_i|^2 \bigg) \\
& + \sum_{i\in{I_\text{H}}}
n_i\log{\frac{x_i}{1-x_i}} \Bigg),
\end{aligned}
\label{eq:trialP_MH}
\end{equation}
with the partition function
\begin{equation} 
\mathnormal{\Xi}_0 =
\prod_{i\in{I_\text{M}} \cup I_\text{H}}
\bigg(\dfrac{\sigma_i\sqrt{k_{\text{B}}T_i m_i}}{\hbar}\bigg)^3
\prod_{i\in{I_\text{H}}}
\dfrac{1}{1-x_i},
\label{eq:trialpartfun_MH}
\end{equation}
where $k_\text{B}$ is the Boltzmann constant and $\hbar$ is the reduced Planck's constant. $m_i$ and $T_i$ denote the atomic mass and absolute local temperature, respectively. $\bar{\bm{q}}_i$, $\sigma_i$, $\bar{\bm{p}}_i$, and $x_i$ are parameters characterizing the trial PDF. Furthermore, it can be deduced that $\bar{\bm{q}}_i$ and $\sigma_i$ represent the mean and standard deviation of $\bm{q}_i$, respectively. Similarly, $\bar{\bm{p}}_i$ and $\sqrt{k_{\text{B}}T_i m_i }$ correspond to the mean and standard deviation of $\bm{p}_i$, respectively. $x_i$, referred to as the H atomic faction, denotes the mean of $n_i$.
 
The statistical behaviors of microscopic variables can be calculated through mean-field approximation applied to the system's free energy. Following straightforward derivations as outlined in Ref.~\cite{sun2017acceleration}, we obtain that $\bar{\bm{p}}_i=0$, while $\bar{\bm{q}}_i$ and $\sigma_i$ can be determined by minimizing the variational Gaussian Helmholtz free energy
\begin{equation} 
\begin{aligned}
\min\limits_{ \{\bar{\bm{q}}\},\{\sigma\}}\mathcal{F} \big( \{\bar{\bm{q}}\},\{\sigma\};\{x\} \big) = 
& \langle{V}\rangle_0
+ \dfrac{3}{2}\sum_{\substack{i\in{I_\text{M}}}} k_{\text{B}}T_i\bigg(\log{\dfrac{\hbar^2}{k_{\text{B}} T_i m_i \sigma_i^2}} -1\bigg) \\
& + \dfrac{3}{2}\sum_{\substack{i\in{I_\text{H}}}} k_{\text{B}}T_i\bigg(\log{\dfrac{\hbar^2}{k_{\text{B}}T_i m_i \sigma_i^2}} + x_i - 2\bigg) \\
& + \sum_{\substack{i\in{I_\text{H}}}} k_{\text{B}}T_i\big(x_i\log{x_i}+(1-x_i)\log{(1-x_i)}\big),
\end{aligned}
\label{eq:fren0}
\end{equation}
where $V\big(\{\bm{q}\}, \{n\} \big)$ is the interatomic potential, and $\langle{\cdot}\rangle_0$ denotes the average computed using the trial PDF $\rho_0$. It's worth noting that in accordance with the Gibbs-Bogoliubov inequality, the Helmholtz free energy estimated by Eq.~(\ref{eq:fren0}) serves as an upper bound on the true Helmholtz free energy, as discussed in Ref.~\cite{venturini2014atomistic}.

\subsection{A discrete kinetic law}

The nonequilibrium thermodynamics model formulated in Eq.~(\ref{eq:fren0}) requires kinetic laws to capture the time evolution of the atomic fields, i.e., atomic fraction $x_i$ and atomic temperature $T_i$. In this work, we focus on the equilibration of the atomic configuration under a constant chemical potential. Therefore, the atomic temperature $T_i$, becomes uniform over all the sites, and is equal to the temperature $T$ defined in equilibrium thermodynamics. Then we can formulate the chemical potential at site $i$ by differentiating Eq.~(\ref{eq:fren0}) with respect to $x_i$, i.e.,
\begin{equation}
\begin{aligned}
\mu_i & = \dfrac{\partial\mathcal{F}}{\partial x_i} \\
& = \dfrac{3}{2} k_\text{B}T + k_\text{B}T\log{\dfrac{x_i}{1-x_i}}+\dfrac{\partial\langle{V}\rangle_0}{\partial{x_i}},\quad i \in I_\text{H},
\end{aligned}
\label{eq:chempot}
\end{equation}
and also define the formation energy at site $i$
\begin{equation}
f_i = \mu_i - k_\text{B}T\log{\dfrac{x_i}{1-x_i}},\quad i \in I_\text{H},
\label{eq:exchange}
\end{equation}
which excludes the contribution of the configurational entropy from the chemical potential.

Within the DMD model, a discrete kinetic law is responsible for driving slow mass transport at the atomistic length scale. At any time point, it enforces the balance of mass at each interstitial site, i.e.,
\begin{equation}
\dot{x}_i = \sum\limits_{j\neq i} J_{ij}, \quad i,j\in{I_\mathrm{H}},
\label{eq:diff}
\end{equation}
where $J_{ij}$ denotes the {\it bondwise} mass flux between site $i$ and site $j$, and satisfies $J_{ij} = -J_{ji}$. Following Refs.~\cite{li2011diffusive, dontsova2014solute, dontsova2015solute, mendez2018diffusive}, we assume that the mass exchange $J_{ij}$ between two sites is governed by the master equation
\begin{equation}
J_{ij} = v \exp \bigg[-\dfrac{Q_{ij}}{k_\text{B}T} \bigg] \bigg\{ x_j(1-x_i) \exp \bigg[ \dfrac{f_{ji}}{2k_\text{B}T} \bigg]  - x_i(1-x_j) \exp \bigg[ \dfrac{f_{ij}}{2k_\text{B}T} \bigg]\bigg\},
\label{eq:master}
\end{equation}
where $v$ is the jump attempt frequency, and $f_{ij}=f_i-f_j$ is the difference of the formation energies between site $i$ and $j$. $Q_{ij}=F^*_{ij}-(f_i+f_j)/2$ is the activation energy barrier for H jump, where $F^*_{ij}$ is the saddle height energy from site $i$ to $j$. The first term in the curly brackets corresponds to the contribution of the mass flux from site $j$ to $i$, whereas the second denotes that from $i$ to $j$. The factor $x_j (1-x_i )$ means that the diffusional H jumps from site $j$ to $i$ can only take place when site $j$ is occupied and site $i$ is vacant. The mass exchange cannot occur kinematically if both sites are fully vacant or fully occupied. Then substituting Eq.~(\ref{eq:exchange}) into Eq.~(\ref{eq:master}) yields
\begin{equation}
J_{ij} = v \exp \bigg[-\dfrac{F^*_{ij}}{k_\text{B}T} \bigg] (1-x_i) (1-x_j)  \bigg\{ \exp \bigg[ \dfrac{\mu_j}{k_\text{B}T} \bigg]  - \exp \bigg[ \dfrac{\mu_i}{k_\text{B}T} \bigg]\bigg\}.
\label{eq:masterchem}
\end{equation}

By examining Eq.~(\ref{eq:masterchem}), it's apparent that the mass flow between interstitial sites $i$ and $j$ is predominantly governed by the gradient of the exponential function of their respective local chemical potentials $\mu_i$ and $\mu_j$. If the difference of the chemical potentials is zero, then these two sites will be in local equilibrium and the net mass exchange between them is zero. This is also consistent with the grand-canonical equilibrium thermodynamics.

\subsection{Equilibration under a constant chemical potential}
\label{sec:equil}

The DMD theory operates within the grand canonical ensemble ($\mu$VT), which permits the exchange of mass between atomic interstitial sites and also between the simulation domain and an infinitely large H reservoir. By integrating the kinetic law described in Eq.~(\ref{eq:masterchem}), H is allowed to flow locally between sites and globally in or out of the system through its boundaries. Subsequently, given a nonequilibrium spatial distribution of H, minimizing the free energy as in Eq.~(\ref{eq:fren0}) enables energetic relaxation of the system. The diffusion of H halts when the chemical potentials, as defined in Eq.~(\ref{eq:chempot}), become uniform across all interstitial sites. This convergence marks the attainment of chemical equilibrium in the system, characterized by a possibly nonuniform distribution of H atoms and the resulting lattice deformation on the host metal lattice.

In this study, our objective is to determine the equilibrium atomic configuration within a H environment characterized by a constant chemical potential $\mu_\text{H}$. To achieve this, we aim to compute the values of $\{\bar{\bm{q}}\}$, $\{\sigma\}$, and $\{x\}$ such that $\mu_i = \mu_\text{H}$ holds for all $i\in{I_\text{H}}$, while ensuring the system attains the lowest free energy $\mathcal{F}$. Rather than directly tackling the time-dependent pairwise diffusion equations outlined in Eqs.~(\ref{eq:diff}) and (\ref{eq:masterchem}), we propose an alternative iteration strategy involving a ``pseudo-diffusion" approach to update the H atomic fraction at each site. In particular, the change in $x_i$ at the $n$-th iteration is determined by
\begin{equation}
\Delta x_i^{(n)} = \dfrac{B}{k_\text{B} T} \big(\mu_\text{H}-\mu_i^{(n)} \big), \quad \forall i\in{I_\text{H}},
\label{eq:pseudodiff}
\end{equation}
where $B$ represents a dimensionless positive coefficient that controls the rate of change of $x_i$. The superscript $(n)$ denotes the outcome at the $n$-th iteration. Notably, when $\mu_\text{H}>\mu_i$, $\Delta x_i$ is positive, indicating an influx of H into the site, and vice versa. $x_i$ remains unchanged if $\mu_\text{H}=\mu_i$, suggesting equilibrium between the site and the H environment. The magnitude of $\Delta x_i$ also depends on the difference in chemical potentials between this site and the H environment. Algorithm~\ref{algo:DMD} provides a pseudocode that implements this approach. It is worth noting that through Eq.~(\ref{eq:pseudodiff}), we can only determine the system's equilibrium configuration. By avoiding the solution of pairwise diffusion equations, we aim to significantly expedite computation time and conduct high-throughput calculations for various chemical potentials. The detailed mechanism of H diffusion in metallic nanocrystals is extensively discussed in Refs.~\cite{sun2019atomistic, sun2018long, sun2017acceleration, sun2024exploring}. 

\begin{algorithm}[!ht]
\caption{Solution of the key equations under a constant H chemical potential.}
\label{algo:DMD}
\algblock{Begin}{End}
\begin{algorithmic}[1]
\BState {\bf Input}: $\{x\}^0$ (initial H fraction), $\mu_\text{H}$ (H chemical potential)
\Begin
\BState \emph{initialization}:
\State $n \leftarrow 0$
\State solve Eq.~(\ref{eq:Hcomp}) at $\{x\}^{(0)}$,
\State $\quad\quad \displaystyle \bar{x}^{(0)} \leftarrow  \dfrac{1}{N_\text{H}} \sum_{i\in{I_\text{H}}}x_i^{(0)}$
\State solve Eq.~(\ref{eq:fren0}) at $\{x\}^{(0)}$,
\State $\quad\quad \displaystyle \min\limits_{ \{\bar{\bm{q}}\},\{\sigma\}}\mathcal{F} \big(\{\bar{\bm{q}}\},\{\sigma\};\{x\}^{(0)} \big) \Rightarrow \{\bar{\bm{q}}\}^{(0)},\{\sigma\}^{(0)}$
\State solve Eq.~(\ref{eq:chempot}) at $\{\bar{\bm{q}}\}^{(0)}$, $\{\sigma\}^{(0)}$, and $\{x\}^{(0)}$,
\State $\quad\quad \displaystyle \mu_i^{(0)} \leftarrow 
\dfrac{3}{2} k_\text{B}T  + k_\text{B}T \log{\dfrac{x_i^{(0)}}{1-x_i^{(0)}}}+\left.\dfrac{\partial\langle{V}\rangle_0}{\partial{x_i}}\right|^{(0)},\quad \forall i \in I_\text{H}$
\BState \emph{iteration loop}:
\While{$\big|\bar{x}^{(n+1)}-\bar{x}^{(n)}\big|>\epsilon$}
\State solve Eq.~(\ref{eq:pseudodiff}) at $\{x\}^{(n)}$ and $\{\mu\}^{(n)}$,
\State $\quad\quad \displaystyle x_i^{(n+1)} \leftarrow x_i^{(n)} + \dfrac{B}{k_\text{B} T} \big(\mu_\text{H}-\mu_i^{(n)} \big), \quad \forall i\in{I_\text{H}}$
\State solve Eq.~(\ref{eq:Hcomp})  at $\{x\}^{(n+1)}$,
\State $\quad\quad \displaystyle \bar{x}^{(n+1)} \leftarrow \dfrac{1}{N_\text{H}} \sum_{i\in{I_\text{H}}}x_i^{(n+1)}$
\State solve Eq.~(\ref{eq:fren0}) at $\{x\}^{(n+1)}$,
\State $\quad\quad \displaystyle \min\limits_{ \{\bar{\bm{q}}\},\{\sigma\}}\mathcal{F} \big(\{\bar{\bm{q}}\},\{\sigma\};\{x\}^{(n+1)}\big)\Rightarrow \{\bar{\bm{q}}\}^{(n+1)},\{\sigma\}^{(n+1)}$
\State solve Eq.~(\ref{eq:chempot}) at $\{\bar{\bm{q}}\}^{(n+1)}$, $\{\sigma\}^{(n+1)}$, and $\{x\}^{(n+1)}$,
\State $\quad\quad  \displaystyle \mu_i^{(n+1)} \leftarrow 
\dfrac{3}{2} k_\text{B}T  + k_\text{B}T \log{\dfrac{x_i^{(n+1)}}{1-x_i^{(n+1)}}}+\left.\dfrac{\partial\langle{V}\rangle_0}{\partial{x_i}}\right|^{(n+1)},\quad \forall i \in I_\text{H}$
\State $n\leftarrow n+1$
\EndWhile
\End
\State {\bf Output}: $\{\bar{\bm{q}}\}^{(n)}$, $\{\sigma\}^{(n)}$, $\{x\}^{(n)}$, $\{\mu\}^{(n)}$, and $\bar{x}^{(n)}$
\end{algorithmic}
\end{algorithm}

\subsection{Simulation setup and implementation details}
\label{sec:setup}

\begin{figure}[!ht]
\centering
\includegraphics[width=0.9\textwidth]{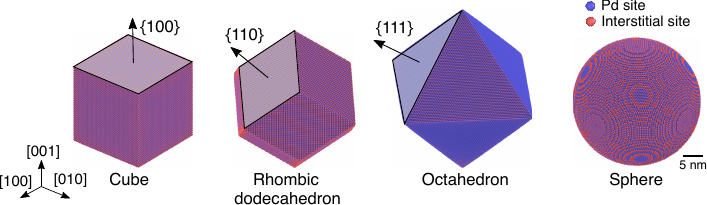}
\caption{Computational models of nanoparticles with a nominal length of $15.6$ nm. The faces of the polyhedra are highlighted by grey planes.}
\label{fig:specimen}
\end{figure}

We investigate the equilibrium configuration during the absorption and desorption of H in Pd nanoparticles of different sizes and shapes at room temperature (i.e., $T=300~\text{K}$). The shapes studied include cube (Cub) with $6$ faces on $\{100\}$ planes, rhombic dodecahedron (Rho) with $12$ faces on $\{110\}$ planes, octahedron (Oct) with $6$ faces on $\{111\}$ planes, and sphere (Sph). These polyhedra are designed to feature crystallographically low-index planes, with each polyhedron having identical faces in terms of shape and size. To ensure reasonable comparisons and quantify size effects, we create a uniform metric across all shapes, termed the nominal length $l$ defined as the cube root of the particle volume. Consequently, particles of the same volume but different shapes are characterized by the same nominal length $l$, equivalent to the length of the cubic particle. We explore four nominal lengths: $3.9$~nm, $7.8$~nm, $11.7$~nm, and $15.6$~nm. Our simulations include both FCC host sites that are occupied by Pd atoms and octahedral interstitial sites within the Pd lattice that can be occupied by H atoms. All particles under consideration are single crystals. Figure~\ref{fig:specimen} illustrates the computational models of the particles with a nominal length of $15.6$ nm. Details regarding the total number of Pd and interstitial sites for each model are provided in Table~\ref{tab:atomnum}.

\begin{table}[!ht]
\centering
\begin{tabular}{|c|c|c|c|}
\hline
Nominal length (nm)   & Shape & Number of Pd sites & Number of interstitial sites \\ \hline
\multirow{4}{*}{3.9}  & Cub   & 4,000              & 4,000                        \\ \cline{2-4} 
                      & Rho   & 3,727              & 3,728                        \\ \cline{2-4} 
                      & Oct   & 4,579              & 3,894                        \\ \cline{2-4} 
                      & Sph   & 3,589              & 3,564                        \\ \hline
\multirow{4}{*}{7.8}  & Cub   & 32,000             & 32,000                       \\ \cline{2-4} 
                      & Rho   & 31,263             & 31,264                       \\ \cline{2-4} 
                      & Oct   & 33,781             & 31,116                       \\ \cline{2-4} 
                      & Sph   & 32,565             & 32,702                       \\ \hline
\multirow{4}{*}{11.7} & Cub   & 108,000            & 108,000                      \\ \cline{2-4} 
                      & Rho   & 107,183            & 107,184                      \\ \cline{2-4} 
                      & Oct   & 110,935            & 104,994                      \\ \cline{2-4} 
                      & Sph   & 105,981            & 106,114                      \\ \hline
\multirow{4}{*}{15.6} & Cub   & 256,000            & 256,000                      \\ \cline{2-4} 
                      & Rho   & 256,063            & 256,064                      \\ \cline{2-4} 
                      & Oct   & 259,369            & 248,856                      \\ \cline{2-4} 
                      & Sph   & 261,563            & 261,742                      \\ \hline
\end{tabular}
\caption{Numbers of Pd and interstitial sites in the numerical simulations.}
\label{tab:atomnum}
\end{table}

To characterize H storage capacity of particles, we define the H composition $\bar{x}$ as
\begin{equation}
\bar{x} = \dfrac{1}{N_\text{H}} \sum_{i\in{I_\text{H}}}x_i.
\label{eq:Hcomp}
\end{equation}
Throughout the simulations, the chemical potential of H environment $\mu_\text{H}$ remains fixed within the range of $-4.5$~eV to $-2$~eV. For H absorption simulations, initial H fraction values are set to $10^{-16}$, and the lattice parameter of the particle is initialized at $a_\text{L}=3.894~\text{\AA}$ for the initial relaxation of the system. By contrast, for H desorption simulations, initial H fraction values are set to $1-10^{-16}$, and the initial lattice parameter is set to $a_\text{L}=4.314~\text{\AA}$. This setup enables the H composition to increase during absorption and decrease during desorption iterations. Regarding the dimensionless coefficient $B$, we have conducted tests using various values ranging from $10^{-1}$ to $10^{-3}$, observing convergence of the equilibrium H distribution as the value increases. In all the simulation results, $B$ is set to $10^{-3}$. The tolerance parameter $\epsilon$ is set to $10^{-6}$. We employ an embedded atom method (EAM) potential~\cite{zhou2008embedded} to capture atom interactions. We minimize the free energy $\mathcal{F}$ through a quasi-Newton Broyden-Fletcher-Goldfarb-Shanno (BFGS) method~\cite{head1985broyden}. The thermodynamic average is calculated by Jensen’s inequality with respect to $\{n\}$ and third-order Gaussian quadratures on a sparse grid with respect to $\{\bm{q}\}$~\cite{sun2017acceleration}. Moreover, we employ OVITO to visualize simulation results and calculate volumetric strains~\cite{stukowski2010visualization}. 

\section{Results and discussions}
\label{sec:result}

\subsection{Absorption and desorption isotherms}
\label{sec:instherm}

\begin{figure}[!ht]
\centering
\includegraphics[width=1.0\textwidth]{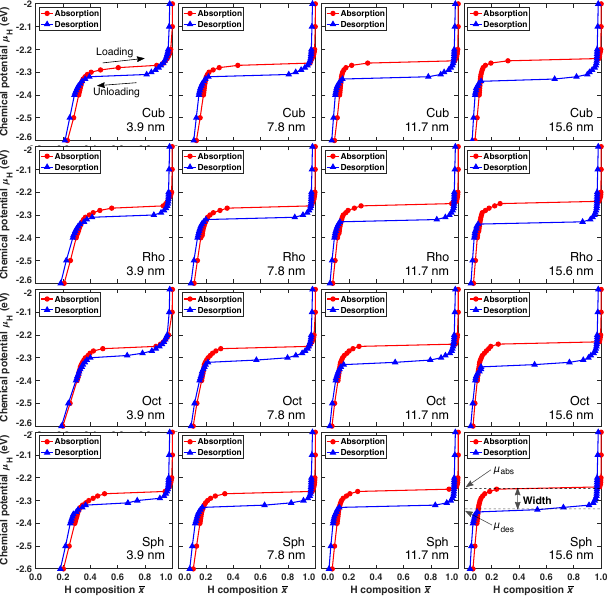}
\caption{Absorption and desorption isotherms of the Pd nanocrystals characterized in this study. The normal lengths of the nanocrystals include $3.9$~nm, $7.8$~nm, $11.7$~nm, and $15.6$~nm. The shapes include cube (Cub), rhombic dodecahedron (Rho), octahedron (Oct), and sphere (Sph). All isotherms are computed at $300$~K. The plateau chemical potentials at absorption and desorption and the hysteresis width are illustrated in the subfigure of sphere with $l=15.6$~nm.}
\label{fig:isotherm}
\end{figure}

We begin our investigation by examining the relationship between the chemical potential $\mu_\text{H}$ and the H composition $\bar{x}$ at equilibrium, commonly referred to as H chemical potential-composition ($\mu_\text{H}$-$\bar{x}$) isotherms. By varying $\mu_\text{H}$, we calculate $\bar{x}$ and construct complete absorption and desorption isotherms for $16$ individual Pd nanocrystals, each with one of the $4$ different sizes and $4$ different shapes listed in Table~\ref{tab:atomnum}. These isotherms are depicted in Fig.~\ref{fig:isotherm}.

Upon analyzing the isotherms as a function of H chemical potential, particle size and shape, several trends are notable. Firstly, most of the particles undergo an abrupt phase transformation from both $\alpha$-to-$\beta$ phase in absorption and $\beta$-to-$\alpha$ phase in desorption, as evidenced by the plateau of the isotherms. Secondly, comparing isotherms of a specific shape reveals that larger particles exhibit more abrupt transitions between the $\alpha$ and $\beta$ phases. This observation contrasts with the sloped transition regions typically observed in ensemble measurements of Pd nanocubes of similar sizes~\cite{bardhan2013uncovering}, suggesting that ensemble results can be averaged by variations in particle size, shape, and surface condition of nanocrystals. Thirdly, the isotherms exhibit distinct hysteresis gaps between the absorption and desorption chemical potentials. These hysteresis gaps widen as particle size increases for all shapes, consistent with experimental measurements of individual Pd nanocubes ranging in size from $13$ nm to $29$ nm~\cite{baldi2014situ}. Fourthly, at $l=3.9$~nm, all particles tend to display sloped transition regions, with shape effects becoming more pronounced due to large surface-area-to-volume ratio.

\begin{figure}[!ht]
\centering
\includegraphics[width=0.7\textwidth]{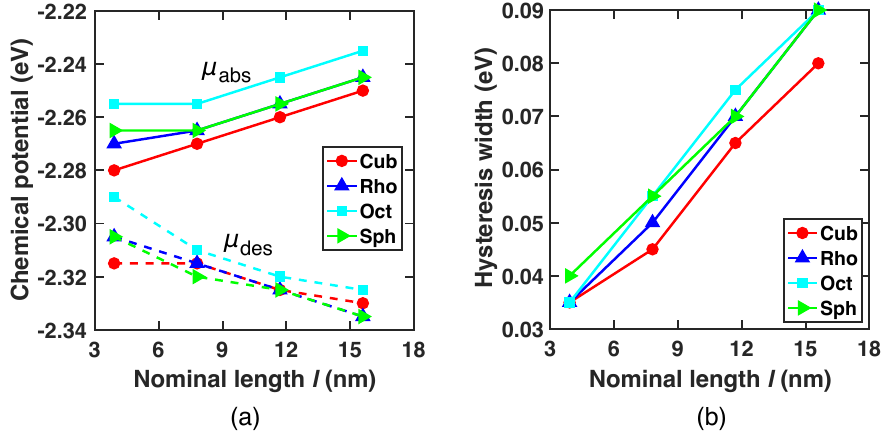}
\caption{Size and shape dependence of sorption hysteresis: (a) Plateau chemical potentials at absorption and desorption, and (b) Hysteresis width between the plateau chemical potentials. In Subfigure (a),  the solid and dashed lines show the chemical potentials at absorption $\mu_\text{abs}$ and desorption $\mu_\text{des}$, respectively.}
\label{fig:hysteresis}
\end{figure}

To quantify the size and shape effects, we calculate the plateau chemical potentials at absorption and desorption, denoted by $\mu_\text{abs}$ and $\mu_\text{des}$, respectively, as well as the hysteresis width for each isotherm. These measures are schematically illustrated in one subfigure of Fig.~\ref{fig:isotherm}. Fig.~\ref{fig:hysteresis} further presents the plateau chemical potentials and hysteresis width for different shapes as a function of the nominal length. From Fig.~\ref{fig:hysteresis}(a), it's evident that for all shapes, the larger particles in our study tend to load H at higher $\mu_\text{abs}$ and unload H at lower $\mu_\text{des}$ compared to smaller ones. Consequently, the hysteresis width, characterized by the difference between $\mu_\text{abs}$ and $\mu_\text{des}$, also increases, as shown in Fig.~\ref{fig:hysteresis}(b). Moreover, Fig.~\ref{fig:hysteresis}(a) highlights some shape-dependent results. Specifically, at a given size, the Cub particle exhibits the lowest $\mu_\text{abs}$, while the Oct particle shows the highest $\mu_\text{abs}$. The $\mu_\text{abs}$ of the Rho and Sph particles are identical, except for $l=3.9$ nm. Regarding desorption, the Oct particle has the highest $\mu_\text{abs}$, whereas the other shapes do not follow a discernible trend. Fig.~\ref{fig:hysteresis}(b) also shows that the Cub samples have the smallest hysteresis widths across all sizes under consideration.

\begin{figure}[!ht]
\centering
\includegraphics[width=1.0\textwidth]{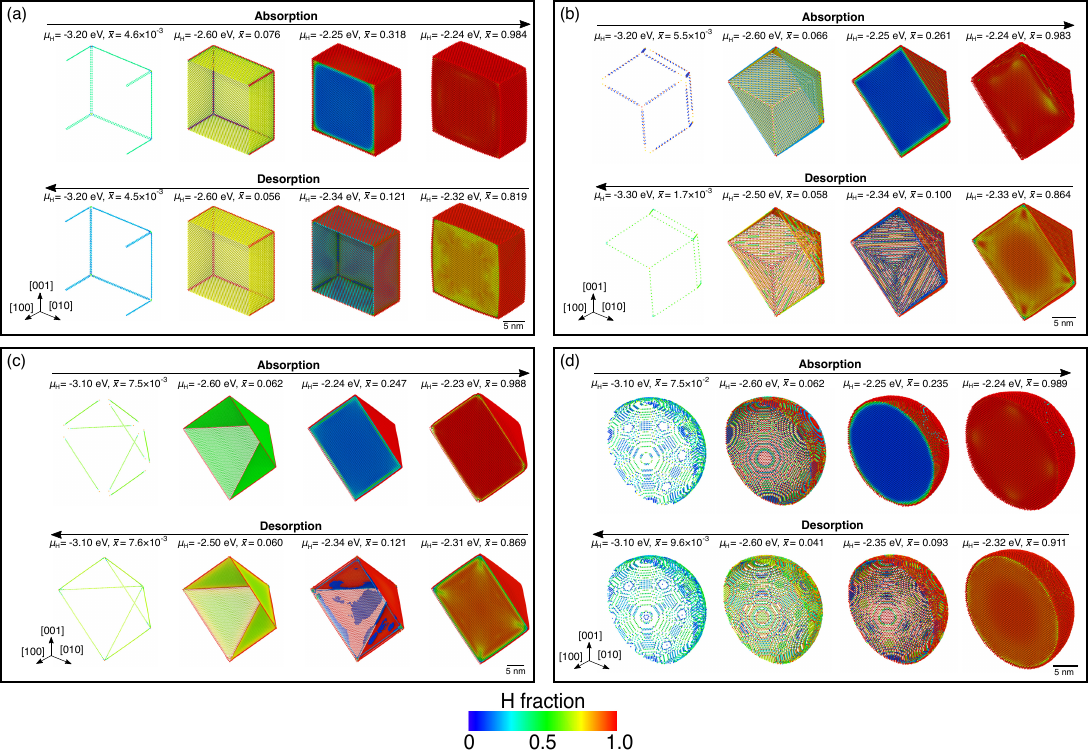}
\caption{H spatial distribution during the absorption and desorption processes for different particle shapes: (a) Cub, (b) Rho, (c) Oct, and (d) Sph. The nominal length for all the particles is $15.6$~nm. Only half of the particles are shown with the $[100]$ cross-section cutting through the middle of the particles. Sites with H fraction less than $0.1$ are removed for the sake of visualization.}
\label{fig:transf}
\end{figure}

\subsection{Hydride and dehydride phase transformation}

We delve deeper into the details of the H loading and unloading processes. Fig.~\ref{fig:transf} illustrates the H spatial distribution during the absorption and desorption processes for the $4$ particles with different shapes but the same nominal length of $15.6$ nm. Notably, for all particles considered, H atoms saturate interstitial sites in the vicinity of the surface of nanocrystalline Pd at $\mu_\text{H}$ well below the plateau chemical potential $\mu_\text{abs}$ due to insufficient atomic coordination near the surface. Consequently, this region near the surface undergoes $\alpha$-to-$\beta$ phase transformation at lower $\mu_\text{H}$ than the bulk. This finding aligns with experimental observations in Pd nanoparticles with sizes around $5$ nm~\cite{sachs2001solubility} and around $20$ nm~\cite{narayan2016reconstructing}. It also confirms the trend of sloped phase transformation when the particle size is small (see Section~\ref{sec:instherm}). Additionally, our simulations reveal that the saturation ordering near the surface of polyhedral particles (i.e., Cub, Rho, and Oct) significantly depends on the coordination number of interstitial sites. Specifically, as $\mu_\text{H}$ increases, H atoms initially saturate sites at the edges with the lowest coordination number, followed by the faces with intermediate coordination, and finally in the bulk with full coordination. This saturation ordering remains consistent across all polyhedral particles with all sizes under consideration. By contrast, the Sph particles do not exhibit clear surface saturation ordering. 

The desorption process can be regarded as a reverse of the absorption process in terms of the locations of dehydriding sites. As $\mu_\text{H}$ decreases during desorption, H atoms are unloaded firstly from the interior of the particles and then the surface. For all the polyhedra considered, the desorption ordering in the surface is first the faces followed by the edges.

\subsection{Surface induced size- and shape-dependence}

\subsubsection{Absorption}

\begin{figure}[!ht]
\centering
\includegraphics[width=1.0\textwidth]{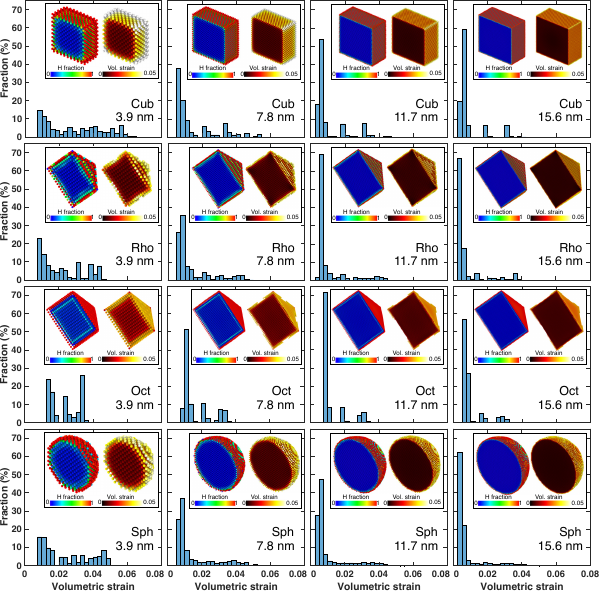}
\caption{Volumetric strain statistics and distribution fields in the $\alpha$ phase at $\mu_\text{H}=-2.29$~eV in absorption. In each inset, the left and right figures depict the distributions of H fractions and volumetric strains for a half of particles, respectively.}
\label{fig:strain_abs}
\end{figure}

\begin{figure}[!ht]
\centering
\includegraphics[width=0.7\textwidth]{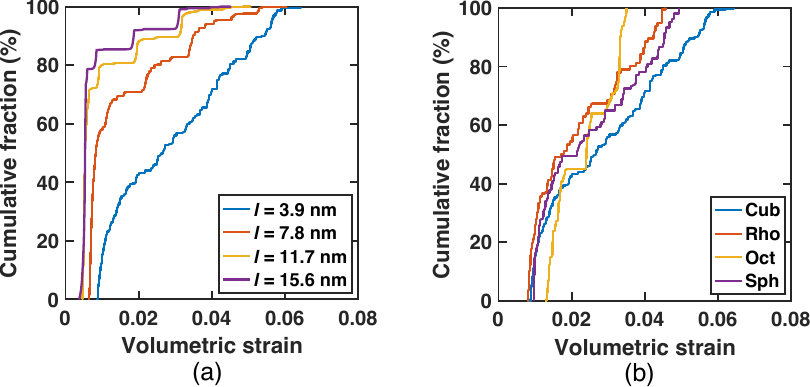}
\caption{Cumulative fraction of volumetric strain for (a) Cub particles with different sizes, and (b) particles with same size $l=3.9$~nm but different shapes. The results are obtained at $\mu_\text{H}=-2.29$~eV in absorption.}
\label{fig:cdf_abs}
\end{figure}

To comprehend the size- and shape-dependent findings discussed earlier, we delve into the lattice deformation at all Pd sites induced by free surface and inhomogeneous H distribution. Specifically, we utilize a perfect and undistorted crystal with a lattice constant of $a_\text{L}=3.894~\text{\AA}$, matching the $\alpha$ phase in our simulations, as a reference to compute volumetric strains at Pd sites in equilibrium atomic configurations during absorption. Fig.~\ref{fig:strain_abs} presents histograms of the volumetric strains at $\mu_\text{H}=-2.29$~eV in absorption, accompanied by insets depicting the distributions of H fractions and volumetric strains. For more quantitative comparisons, Fig.~\ref{fig:cdf_abs} shows two examples of cumulative fractions at $\mu_\text{H}=-2.29$~eV in absorption, obtained by computing the proportion of sites with volumetric strains less than a given value.

Notably, the insets of Fig.~\ref{fig:strain_abs} reveal that all particles exhibit a high H concentration shell surrounding a low H concentration core, given that $\mu_\text{H}$ is well below $\mu_\text{abs}$ for all particles. Consequently, all sites undergo volumetric expansion, with the expansion in the shell significantly larger than that in the core due to the lattice expansion from $\alpha$ to $\beta$ phase. Furthermore, the histograms illustrate that for a given shape, as the particle size increases, the proportion of sites with large strains decreases. This trend can be further clarified by the cumulative fraction for the Cub particles with varying sizes, as shown in Fig.~\ref{fig:cdf_abs}(a). For instance, the proportion of sites with strains larger than $0.04$ is $29\%$ for $l=3.9$~nm, $6\%$ for $7.8$~nm, $1\%$ for $11.7$~nm, and $0$ for $15.6$~nm. Consequently, smaller particles tend to provide more space in the H-poor Pd core for H atoms to occupy, facilitating H loading at low chemical potentials. This observation may explain why smaller particles exhibit lower $\mu_\text{abs}$ than larger ones, as discussed in Section~\ref{sec:instherm}.

The shape effects can be explained by a similar manner. For a given size, the Cub sample has the highest proportion of highly expanded sites, leading to the lowest loading $\mu_\text{abs}$. By contrast, the Oct sample exhibits the least proportion of highly expanded sites, resulting in the highest $\mu_\text{abs}$ among the $4$ shapes. To further clarify this finding, the cumulative fraction for the size of $l=3.9$~nm is shown in Fig.~\ref{fig:cdf_abs}(b). For example, the proportion of sites with strains larger than $0.04$ is $29\%$ for Cub, $22\%$ for Sph, $11\%$ for Rho, and $0$ for Oct.

\subsubsection{Desorption}

\begin{figure}[!ht]
\centering
\includegraphics[width=1.0\textwidth]{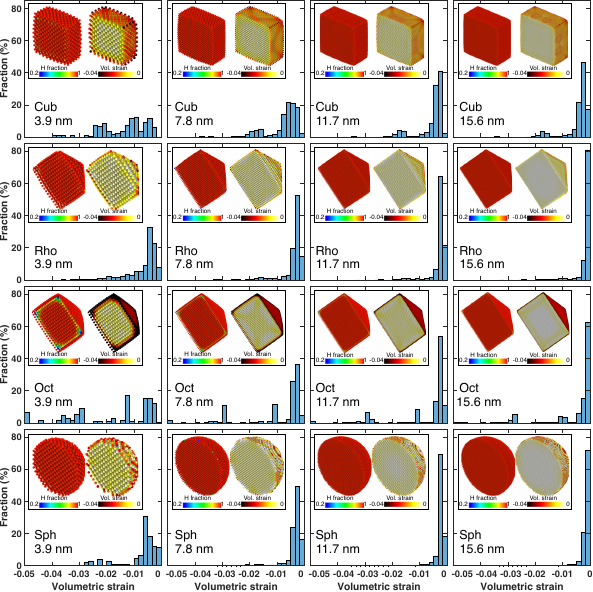}
\caption{Volumetric strain statistics and distribution fields in the $\beta$ phase at $\mu_\text{H}=-2.21$~eV in desorption. In each inset, the left and right figures depict the distributions of H fractions and volumetric strains for a half of particles, respectively.}
\label{fig:strain_des}
\end{figure}

\begin{figure}[!ht]
\centering
\includegraphics[width=0.7\textwidth]{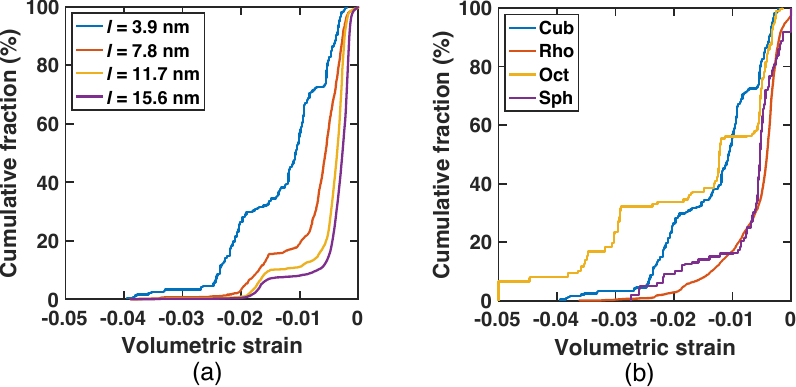}
\caption{Cumulative fraction of volumetric strain for (a) Cub particles with different sizes, and (b) particles with same size $l=3.9$~nm but different shapes. The results are obtained at $\mu_\text{H}=-2.21$~eV in desorption.}
\label{fig:cdf_des}
\end{figure}

Fig.~\ref{fig:strain_des} presents histograms of the volumetric strains at $\mu_\text{H}=-2.21$ eV in desorption, with insets also showing the distributions of H fractions and volumetric strains in each histogram. These volumetric strains are calculated using a perfect lattice with $a_\text{L}=4.314~\text{\AA}$ as a reference, corresponding to the $\beta$ phase in our simulations. Additionally, Fig.~\ref{fig:cdf_des} illustrates two examples of cumulative fractions at $\mu_\text{H}=-2.21$~eV in desorption.

As evident from the insets of Fig.~\ref{fig:strain_des}, all particles feature both a high H concentration shell and core since $\mu_\text{H}=-2.21$~eV is higher than $\mu_\text{des}$ for all particles, with no $\beta$-to-$\alpha$ transition occurring in this case. All sites undergo compression, with the volumetric strain in the shell significantly smaller than that in the core due to surface stresses. Since compression results in negative strain values, smaller strains indicate more significant compression. Therefore, the lattices in the vicinity of the surface are highly compressed compared to those in the core. Furthermore, the histograms illustrate that for a given shape, as particle size increases, the proportion of sites with small strains decreases. This trend is further clarified by the cumulative fraction for Cub particles with varying sizes, as shown in Fig.~\ref{fig:cdf_des}(a). For instance, the proportion of sites with strains smaller than $-0.02$ is $26\%$ for $l=3.9$ nm, $3\%$ for $7.8$ nm, $0.9\%$ for $11.7$ nm, and $0.4\%$ for $15.6$ nm. These highly compressed spaces tend to expel H atoms from the crystal, facilitating H unloading at high chemical potentials. Consequently, smaller particles exhibit higher $\mu_\text{des}$ during the $\beta$-to-$\alpha$ transformation than larger ones, as discussed in Section~\ref{sec:instherm}.

Similarly, shape effects on $\mu_\text{des}$ can be explained. For a given size, the Oct sample has the highest proportion of highly compressed sites, leading to the highest unloading $\mu_\text{des}$. To clarify this finding, the cumulative fraction for a size of $l=3.9$ nm is shown in Fig.~\ref{fig:cdf_des}(b). For instance, the proportion of sites with strains smaller than $-0.03$ is $23\%$ for Oct, $3\%$ for Cub, $0.4\%$ for Rho, and $0$ for Sph.

\section{Concluding remarks}
\label{sec:concl}

We have conducted a comprehensive analysis of hydrogenation and dehydrogenation of Pd nanoparticles, with different sizes ranging from $3.9$ nm to $15.6$ nm and various shapes including cube, rhombic dodecahedron, octahedron, and sphere. Our methodology combines the DMD approach with an iteration strategy, aiming to minimize the system's free energy and ensure uniform chemical potential consistent with a surrounding H environment. Through numerical simulations, we have explored absorption and desorption isotherms, plateau chemical potentials, hysteresis gaps, and saturation ordering in hydride and dehydride phase transformations. Atomic volumetric strains have been utilized to elucidate the underlying size and shape effects.

Our calculations unveil several noteworthy findings. Firstly, we have observed an abrupt phase transformation characterized by a flat plateau in the chemical potential-composition isotherms across all examined particles during both H loading and unloading processes. Furthermore, a distinct hysteresis gap is evident between absorption and desorption chemical potentials. Notably, as particle size increases, the plateau flattens, and absorption chemical potential increases while desorption chemical potential decreases, thereby widening the hysteresis gap across all shapes. Concerning absorption chemical potential, the size effect is caused by the proportion of highly expended lattice spaces in the $\alpha$ phase core induced by inhomogeneous H distribution before hydride phase transformation. As particle size increases, this proportion decreases. On the other hand, for desorption chemical potential, the size effect arises from different proportion of highly compressed lattice spaces in the $\beta$ phase core induced by surface stress before dehydride phase transformation. As particle size increases, this proportion also decreases.

Regarding shape effects, cubic particles exhibit the lowest absorption chemical potentials at a given size, while octahedral particles demonstrate the highest. This distinction arises from cubic particles with $\{100\}$-oriented faces offering the highest proportion of highly expanded lattice spaces, whereas octahedral particles with $\{111\}$-oriented faces provided the lowest. Octahedral particles also exhibit the highest desorption chemical potentials, attributed to the highest proportion of highly compressed spaces before H unloading induced by $\{111\}$-oriented faces. Furthermore, for all particles considered, H atoms saturate interstitial sites near the surface of Pd nanoparticles at chemical potentials well below the transition chemical potential. Moreover, all polyhedral particles exhibit the same ordering of saturation locations in absorption, starting from edges, followed by faces, and finally the bulk. On the other hand, the desorption process can be treated as a reverse of  the absorption process in terms of dehydriding locations.

The present study primarily focuses on size and shape effects on the equilibrium properties of nanosized crystalline Pd-H systems such as H storage capacity and thermodynamics. However, different morphologies may also affect kinetic properties, such H charging and discharging time and rate. For instance, experiments based on ensemble measurements have unveiled that the hydrogenation speed of Pd $\{111\}$-terminated nanooctahedra is faster than that of $\{100\}$-terminated nanocubes~\cite{li2014shape}. However, it's worth noting that the nanooctahedra considered therein may have a smaller volume than the nanocubes, which could also impact the hydrogenation speed. For a given volume, DMD simulations have suggested that cubic particles aligned with $\{100\}$ planes may be optimal for H storage applications compared to octahedral and spherical particles~\cite{sun2019atomistic}. This superiority arises from the suppression of stacking fault formation and the persistent coherence of the phase boundary in the nanocubes. On the contrary, experiments based on individual measurements have indicated that both $\{100\}$-terminated nanocubes and $\{111\}$-terminated nanooctahedra take approximately the same time to reach equilibrium~\cite{sytwu2018visualizing}. This finding may also neglect size effects. These discrepancies indicate that the size, shape and even surface conditions such as defective or crystalline may affect the sorption kinetics simultaneously. Therefore, these and other topics suggest themselves as worthwhile directions for further studies.

\section*{Acknowledgments}

The material is based upon work supported by NASA Kentucky under NASA award No: 80NSSC20M0047. XS gratefully acknowledges the support from the University of Kentucky through the faculty startup fund and the e-RPA seed grant program. We would thank the University of Kentucky Center for Computational Sciences and Information Technology Services Research Computing for their support and use of the Lipscomb Compute Cluster and associated research computing resources.

\bibliography{mybibfile}
\bibliographystyle{unsrt}

\end{document}